\documentclass{amsart}
\usepackage{amsmath,amssymb,amsfonts,bm}

\usepackage[english]{babel}
\usepackage{color}
\usepackage{tikz}
\usepackage{pgfplots}
\pgfplotsset{compat=newest}
\pgfplotsset{width=7cm, compat=1.10}

\usetikzlibrary{decorations.markings}

\usepgfplotslibrary{fillbetween}
\pgfmathdeclarefunction{poly}{0}{\pgfmathparse{20*x*(x-1)*(x+1)}}
\pgfmathdeclarefunction{poly2}{0}{\pgfmathparse{-20*x*(x-1)*(x+1)}}
\pgfmathdeclarefunction{upperpar}{0}{\pgfmathparse{20*(x^2+0.2)}}
\pgfmathdeclarefunction{lowerpar}{0}{\pgfmathparse{20*(-x^2-0.2)}}

\newcommand{\verteq}{\rotatebox{90}{$\,=$}}

\usepackage{mathtools}

\title{Three quick recipes with fully extended oriented 2d TQFTs}

\author{Domenico Fiorenza}
\address{Sapienza Universit\`a di Roma; Dipartimento di Matematica ``Guido Castelnuovo'', P.le Aldo Moro, 5 - 00185 - Roma, Italy}
\email{fiorenza@mat.uniroma1.it}
\begin{document}

\begin{abstract}
It always happens: you have a talk for dinner and nothing prepared. Your signature dish never fails, but you have served it too many times already and you'd like to surprise your guests with something new. Try these quick, light and colourful reinterpretations of haute cuisine classics, like (nonabelian) Fourier transforms and the Plancherel theorem for finite groups. 
\end{abstract}
	\maketitle
	
\setcounter{tocdepth}{1}	
\tableofcontents

\section{The ingredients}
First thing you need to become familiar with are the ingredients you are going to cook. All of them can be easily obtained in pre-cooked form from the literature in a wide variety of examples. Yet, as the best bread is the one from one's own oven, we'll see how to prepare a few of these out of raw ingredients such as finite groups.

\begin{itemize}
\item  a symmetric monoidal $(\infty,2)$-category $\mathcal{C}$;\footnote{There was a time when the newspapers said that only twelve men\footnotemark \,\!\! understood symmetric monoidal $(\infty,1)$-categories. I do not believe there ever was such a time. There might have been a time when only one man did, because he was the only guy who caught on, before he wrote his paper. But after people read the paper a lot of people kind of understood the theory of symmetric monoidal $(\infty,1)$-categories in some way or other, but more than twelve. On the other hand, I think I can safely say that nobody understands symmetric monoidal $(\infty,2)$-categories.}
\footnotetext{The use of the masculine ``men'' for ``people'' witnesses the acute gender bias at the time of the original quote (Richard Feynman, Cornell University Lectures, 1964).} 

\item a pair of fully dualizable objects in $\mathcal{C}$: the blue object $\color{blue}B$ and the red object $\color{red}R$, together with an \emph{invertible} oriented defect line between them;
\item the \emph{unit} object $\mathbf{1}$ of $\mathcal{C}$, corresponding to the invisible colour;
\item boundary conditions $L,M,N,\dots,$ for the blue and the red object ((possibly) noninvertible defect lines between the unit object and the blue/red object or vice versa);
\item defect points $\phi:L\to M$, $\psi:N\to N$, \dots, on the boundaries. 
\end{itemize}
All these data are required to be homotopy invariant for the $SO(2)$-action on $\mathcal{C}$.

\section{Basic preparations}
In order to be able to prepare a recipe with the above ingredients, you need become familiar with a few basic preparations with fancy names. Notice that only a few of these preparations actually need the defect line to be invertible.
\vskip .4 cm
\noindent{\it Zorro moves:}

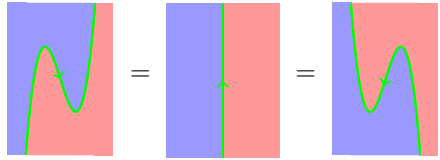
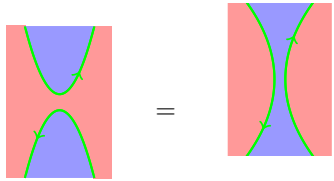
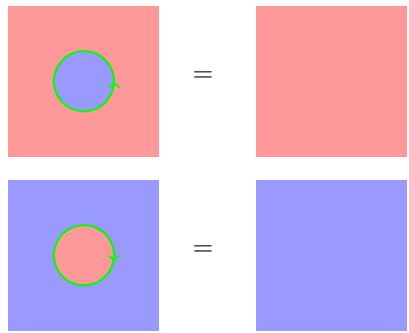
\[
\raisebox{-1.2cm}{
\begin{tikzpicture}
\fill[blue!40!white] (0,0.2) rectangle (0.255,2.226);
\fill[red!40!white] (1.15,0.197) rectangle (1.4,2.226);
  \begin{axis}[hide axis,
 width= 3cm,
height=4cm,
    samples     = 160,
    domain      = -1.3:1.3,
    xmin = -2, xmax = 2,
  ]
  \addplot[name path=poly, green, thick, mark=none, ] {poly};
  \addplot[name path=line, blue!40, no markers, line width=1pt] {17.8};
  \addplot[name path=linetwo, red!40, no markers, line width=1pt] {-17.7};
  \addplot[blue!40] fill between[
    of = poly and line, 
      ];
       \addplot[red!40] fill between[
    of = linetwo and poly, 
      ];
   \addplot[name path=poly, green, thick, postaction={decorate, decoration={markings,
mark=at position 0.5 with {\arrow{<};}}},] {poly};
\end{axis}
\end{tikzpicture}
}
=
\raisebox{-1.05cm}{
\begin{tikzpicture}
\fill[blue!40!white] (-.75,0) rectangle (0,2.05);
\fill[red!40!white] (0,0) rectangle (.75,2.05);
\draw[green,thick,decoration={markings, mark=at position 0.5 with {\arrow{<}}},
        postaction={decorate}]  (0,0) -- (0,2.05);
\end{tikzpicture}
}
=
\raisebox{-1.2cm}{
\begin{tikzpicture}
\fill[blue!40!white] (0,0.2) rectangle (0.255,2.226);
\fill[red!40!white] (1.15,0.197) rectangle (1.4,2.226);
  \begin{axis}[hide axis,
 width= 3cm,
height=4cm,
    samples     = 160,
    domain      = -1.3:1.3,
    xmin = -2, xmax = 2,
  ]
  \addplot[name path=poly2, green, thick, mark=none, ] {poly2};
  \addplot[name path=line, red!40, no markers, line width=1pt] {17.8};
  \addplot[name path=linetwo, blue!40, no markers, line width=1pt] {-17.7};
  \addplot[red!40] fill between[
    of = poly2 and line, 
      ];
       \addplot[blue!40] fill between[
    of = linetwo and poly2, 
      ];
   \addplot[name path=poly2, green, thick, postaction={decorate, decoration={markings,
mark=at position 0.5 with {\arrow{>};}}},] {poly2};
\end{axis}
\end{tikzpicture}
}
\]

\vskip .8 cm 

\noindent {\it Pinch until split:}
\[
\raisebox{-1cm}{
\begin{tikzpicture}
\fill[red!40!white] (0,0.2) rectangle (0.255,2.226);
\fill[red!40!white] (1.15,0.197) rectangle (1.4,2.226);
  \begin{axis}[hide axis,
 width= 3cm,
height=4cm,
    samples     = 160,
    domain      = -1.3:1.3,
    xmin = -2, xmax = 2,
  ]
  \addplot[name path=upperpar, green, thick, mark=none, ] {upperpar};
   \addplot[name path=lowerpar, green, thick, mark=none, ] {lowerpar};
  \addplot[name path=line, blue!40, no markers, line width=1pt] {37.35};
  \addplot[name path=linetwo, blue!40, no markers, line width=1pt] {-37.15};
  \addplot[blue!40] fill between[
    of = upperpar and linetwo, 
      ];
            \addplot[blue!40] fill between[
    of = line and lowerpar, 
      ];
        \addplot[red!40] fill between[
    of = lowerpar and upperpar, 
      ];
   \addplot[name path=upperpar, green, thick, postaction={decorate, decoration={markings,
mark=at position 0.7 with {\arrow{<};}}},] {upperpar};
 \addplot[name path=lowerpar, green, thick, postaction={decorate, decoration={markings,
mark=at position 0.3 with {\arrow{>};}}},] {lowerpar};
\end{axis}
\end{tikzpicture}
}
\quad=\quad 
\raisebox{=-0.915cm}{
\rotatebox{90}{
\begin{tikzpicture}
  \begin{axis}[hide axis,
 width= 3.609cm,
height=3.2cm,
    samples     = 160,
    domain      = -1.3:1.3,
    xmin = -1, xmax = 1,
  ]
  \addplot[name path=upperpar, green, thick, mark=none, ] {upperpar};
   \addplot[name path=lowerpar, green, thick, mark=none, ] {lowerpar};
  \addplot[name path=line, red!40, no markers, line width=1pt] {37.35};
  \addplot[name path=linetwo, red!40, no markers, line width=1pt] {-37.15};
  \addplot[red!40] fill between[
    of = upperpar and linetwo, 
      ];
            \addplot[red!40] fill between[
    of = line and lowerpar, 
      ];
        \addplot[blue!40] fill between[
    of = lowerpar and upperpar, 
      ];
   \addplot[name path=upperpar, green, thick, postaction={decorate, decoration={markings,
mark=at position 0.3 with {\arrow{>};}}},] {upperpar};
 \addplot[name path=lowerpar, green, thick, postaction={decorate, decoration={markings,
mark=at position 0.7 with {\arrow{<};}}},] {lowerpar};
\end{axis}
\end{tikzpicture}
}}
\]

\vskip .8 cm 

\noindent {\it Bubble (dis)appearances:}

\[
\raisebox{-1cm}{
\begin{tikzpicture}
\fill[red!40!white] (0,0) rectangle (2,2);
\fill[blue!40!white] (1,1) circle (0.4cm);
\draw[green,thick,decoration={markings, mark=at position 1 with {\arrow{<}}},
        postaction={decorate}] (1,1) circle (.4cm);
\end{tikzpicture}}
\quad =\quad 
\raisebox{-1cm}{
\begin{tikzpicture}
\fill[red!40!white] (0,0) rectangle (2,2);
\end{tikzpicture}}
\phantom{mmm}
\]
\[
\raisebox{-1cm}{
\begin{tikzpicture}
\fill[blue!40!white] (0,0) rectangle (2,2);
\fill[red!40!white] (1,1) circle (0.4cm);
\draw[green,thick,decoration={markings, mark=at position 1 with {\arrow{>}}},
        postaction={decorate}] (1,1) circle (.4cm);
\end{tikzpicture}}
\quad =\quad 
\raisebox{-1cm}{
\begin{tikzpicture}
\fill[blue!40!white] (0,0) rectangle (2,2);
\end{tikzpicture}}
\phantom{mmm}
\]

\vskip .8 cm 
\noindent {\it Pushing the defect line to the boundary:}
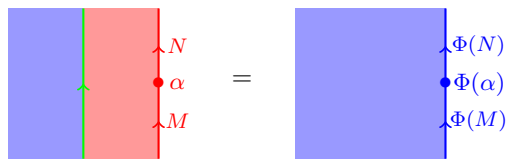
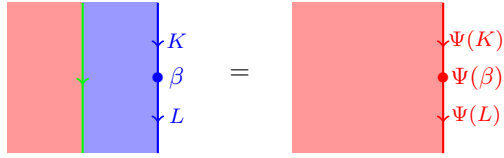
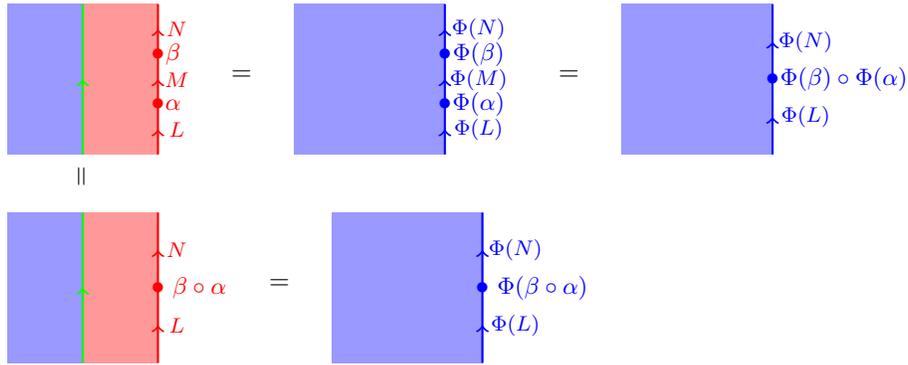
\[
\raisebox{-1cm}{
\begin{tikzpicture}
\fill[blue!40!white] (-1,0) rectangle (0,2);
\fill[red!40!white] (0,0) rectangle (1,2);
\draw[green,thick,decoration={markings, mark=at position 0.5 with {\arrow{<}}},
        postaction={decorate}]  (0,0) -- (0,2);
\draw[red,thick,decoration={markings, mark=at position 0.5 with {\arrow{<}}},
        postaction={decorate}]  (1,1) -- (1,2);
\draw[red,thick,decoration={markings, mark=at position 0.5 with {\arrow{<}}},
        postaction={decorate}]   (1,0) -- (1,1);
\node[red] at (1,1){\textbullet};        
\node[red] at (1.25,1.5){\footnotesize{$N$}};
\node[red] at (1.25,0.5){\footnotesize{$M$}};
\node[red] at (1.25,1){\small{$\alpha$}};
\end{tikzpicture}}
\quad =\quad
\raisebox{-1cm}{
\begin{tikzpicture}
\fill[blue!40!white] (-1,0) rectangle (1,2);
\draw[blue,thick,decoration={markings, mark=at position 0.5 with {\arrow{<}}},
        postaction={decorate}]  (1,1) -- (1,2);
\draw[blue,thick,decoration={markings, mark=at position 0.5 with {\arrow{<}}},
        postaction={decorate}]   (1,0) -- (1,1);
\node[blue] at (1,1){\textbullet};        
\node[blue] at (1.44,1.5){\footnotesize{$\Phi(N)$}};
\node[blue] at (1.44,0.5){\footnotesize{$\Phi(M)$}};
\node[blue] at (1.45,1){\small{$\Phi(\alpha)$}};
\end{tikzpicture}
}
\]
\[
\raisebox{-1cm}{
\begin{tikzpicture}
\fill[red!40!white] (-1,0) rectangle (0,2);
\fill[blue!40!white] (0,0) rectangle (1,2);
\draw[green,thick,decoration={markings, mark=at position 0.5 with {\arrow{>}}},
        postaction={decorate}]  (0,0) -- (0,2);
\draw[blue,thick,decoration={markings, mark=at position 0.5 with {\arrow{>}}},
        postaction={decorate}]  (1,1) -- (1,2);
\draw[blue,thick,decoration={markings, mark=at position 0.5 with {\arrow{>}}},
        postaction={decorate}]   (1,0) -- (1,1);
\node[blue] at (1,1){\textbullet};        
\node[blue] at (1.25,1.5){\footnotesize{$K$}};
\node[blue] at (1.25,0.5){\footnotesize{$L$}};
\node[blue] at (1.25,1){\small{$\beta$}};
\end{tikzpicture}}
\quad = \quad
\raisebox{-1cm}{
\begin{tikzpicture}
\fill[red!40!white] (-1,0) rectangle (1,2);
\draw[red,thick,decoration={markings, mark=at position 0.5 with {\arrow{>}}},
        postaction={decorate}]  (1,1) -- (1,2);
\draw[red,thick,decoration={markings, mark=at position 0.5 with {\arrow{>}}},
        postaction={decorate}]   (1,0) -- (1,1);
\node[red] at (1,1){\textbullet};        
\node[red] at (1.44,1.5){\footnotesize{$\Psi(K)$}};
\node[red] at (1.44,0.5){\footnotesize{$\Psi(L)$}};
\node[red] at (1.45,1){\small{$\Psi(\beta)$}};
\end{tikzpicture}
}
\]

\vskip .8 cm 
\noindent {\it Pushing and glueing:}
\begin{align*}
&\raisebox{-1cm}{
\begin{tikzpicture}
\fill[blue!40!white] (-1,0) rectangle (0,2);
\fill[red!40!white] (0,0) rectangle (1,2);
\draw[green,thick,decoration={markings, mark=at position 0.5 with {\arrow{<}}},
        postaction={decorate}]  (0,0) -- (0,2);
\draw[red,thick,decoration={markings, mark=at position 1/6 with {\arrow{<}} , mark=at position 0.5 with {\arrow{<}} , mark=at position 5/6 with {\arrow{<}} },
        postaction={decorate}]  (1,0) -- (1,2);
\node[red] at (1,2/3){\textbullet}; 
\node[red] at (1,4/3){\textbullet};        
\node[red] at (1.25,1/3){\footnotesize{$L$}};
\node[red] at (1.25,1){\footnotesize{$M$}};
\node[red] at (1.25,5/3){\footnotesize{$N$}};
\node[red] at (1.2,2/3){\small{$\alpha$}};
\node[red] at (1.2,4/3){\small{$\beta$}};
\end{tikzpicture}}
\quad =\quad
\raisebox{-1cm}{
\begin{tikzpicture}
\fill[blue!40!white] (-1,0) rectangle (1,2);
\draw[blue,thick,decoration={markings, mark=at position 1/6 with {\arrow{<}} , mark=at position 0.5 with {\arrow{<}} , mark=at position 5/6 with {\arrow{<}} },
        postaction={decorate}]  (1,0) -- (1,2);
\node[blue] at (1,2/3){\textbullet}; 
\node[blue] at (1,4/3){\textbullet};        
\node[blue] at (1.45,1/3){\footnotesize{$\Phi(L)$}};
\node[blue] at (1.45,1){\footnotesize{$\Phi(M)$}};
\node[blue] at (1.45,5/3){\footnotesize{$\Phi(N)$}};
\node[blue] at (1.45,2/3){\small{$\Phi(\alpha)$}};
\node[blue] at (1.45,4/3){\small{$\Phi(\beta)$}};
\end{tikzpicture}
}
\quad =\quad
\raisebox{-1cm}{
\begin{tikzpicture}
\fill[blue!40!white] (-1,0) rectangle (1,2);
\draw[blue,thick,decoration={markings, mark=at position 0.5 with {\arrow{<}}},
        postaction={decorate}]  (1,1) -- (1,2);
\draw[blue,thick,decoration={markings, mark=at position 0.5 with {\arrow{<}}},
        postaction={decorate}]   (1,0) -- (1,1);
\node[blue] at (1,1){\textbullet};        
\node[blue] at (1.44,1.5){\footnotesize{$\Phi(N)$}};
\node[blue] at (1.44,0.5){\footnotesize{$\Phi(L)$}};
\node[blue] at (1.95,1){\small{$\Phi(\alpha)\circ \Phi(\beta)$}};
\end{tikzpicture}
}
\\
&\qquad\quad \verteq
\\
&\raisebox{-1cm}{
\begin{tikzpicture}
\fill[blue!40!white] (-1,0) rectangle (0,2);
\fill[red!40!white] (0,0) rectangle (1,2);
\draw[green,thick,decoration={markings, mark=at position 0.5 with {\arrow{<}}},
        postaction={decorate}]  (0,0) -- (0,2);
\draw[red,thick,decoration={markings, mark=at position 0.5 with {\arrow{<}}},
        postaction={decorate}]  (1,1) -- (1,2);
\draw[red,thick,decoration={markings, mark=at position 0.5 with {\arrow{<}}},
        postaction={decorate}]   (1,0) -- (1,1);
\node[red] at (1,1){\textbullet};        
\node[red] at (1.25,1.5){\footnotesize{$N$}};
\node[red] at (1.25,0.5){\footnotesize{$L$}};
\node[red] at (1.55,1){\small{$\alpha\circ \beta$}};
\end{tikzpicture}}
\quad =\quad
\raisebox{-1cm}{
\begin{tikzpicture}
\fill[blue!40!white] (-1,0) rectangle (1,2);
\draw[blue,thick,decoration={markings, mark=at position 0.5 with {\arrow{<}}},
        postaction={decorate}]  (1,1) -- (1,2);
\draw[blue,thick,decoration={markings, mark=at position 0.5 with {\arrow{<}}},
        postaction={decorate}]   (1,0) -- (1,1);
\node[blue] at (1,1){\textbullet};        
\node[blue] at (1.44,1.5){\footnotesize{$\Phi(N)$}};
\node[blue] at (1.44,0.5){\footnotesize{$\Phi(L)$}};
\node[blue] at (1.80,1){\small{$\Phi(\alpha\circ \beta)$}};
\end{tikzpicture}
}
\end{align*}
and (of course) the same for blue and red exchanged.

\section{Amuse-bouche: the isomorphism between the centres}
\noindent
{\it Ingredients:} as this is a very simple recipe, you will only need the general ingredients listed at the beginning of the cookbook ($\to$ {\it Ingredients}).
\\ \phantom{i}\\
{\it Difficulty:} easy.
\\ \phantom{i}\\
{\it Cooking time:} 3 minutes.
\\ \phantom{i}\\
{\it Preparation:} First, you need to cook the centres:\footnote{These are called centres as, when $A$ is a semisimple symmetric Frobenius algebra, seen as a fully dualizable and $SO(2)$-homotopy fixed point object in the the symmetric monoidal $(\infty,2)$-category $\mathrm{Alg_2}$ of finite dimensional $\mathbb{K}$-algebras, bimodules and morphisms of bimodules, then $Z(A)$ is precisely the centre of the algebra $A$.}
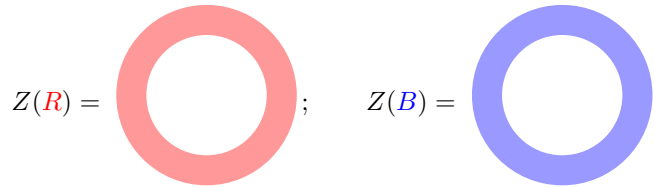
\[
Z({\color{red}R})=\raisebox{-1cm}{
\begin{tikzpicture}
\fill[red!40!white] (0,0) circle (1.2cm);
\fill[white] (0,0) circle (0.8cm);
\end{tikzpicture}}
\,;\qquad Z({\color{blue}B})=\raisebox{-1cm}{
\begin{tikzpicture}
\fill[blue!40!white] (0,0) circle (1.2cm);
\fill[white] (0,0) circle (0.8cm);
\end{tikzpicture}}
\]
Now, you have to prepare isomorphisms between these. To do so, first prepare natural morphisms between them using the invertible defect line:
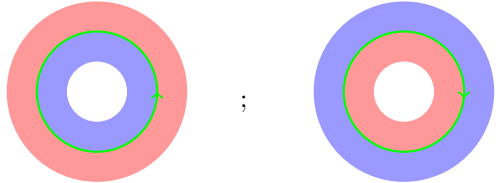
\[
\raisebox{-1cm}{
\begin{tikzpicture}
\fill[red!40!white] (0,0) circle (1.2cm);
\fill[blue!40!white] (0,0) circle (0.8cm);
\fill[white] (0,0) circle (0.4cm);
\draw[green,thick,decoration={markings, mark=at position 1 with {\arrow{<}}},
        postaction={decorate}] (0,0) circle (.8cm);
\end{tikzpicture}}
\qquad ;\qquad 
\raisebox{-1cm}{
\begin{tikzpicture}
\fill[blue!40!white] (0,0) circle (1.2cm);
\fill[red!40!white] (0,0) circle (0.8cm);
\fill[white] (0,0) circle (0.4cm);
\draw[green,thick,decoration={markings, mark=at position 1 with {\arrow{>}}},
        postaction={decorate}] (0,0) circle (.8cm);
\end{tikzpicture}}
\]
Now show that these are indeed isomorphisms (after all, your defect line is invertible, isn't it?). I'll show you only one composition, the other one is identical. First cook the composition
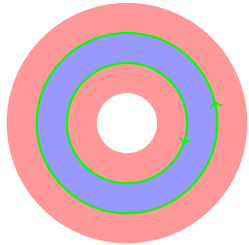
\[
\begin{tikzpicture}
\fill[red!40!white] (0,0) circle (1.6cm);
\fill[blue!40!white] (0,0) circle (1.2cm);
\fill[red!40!white] (0,0) circle (0.8cm);
\fill[white] (0,0) circle (0.4cm);
\draw[green,thick,decoration={markings, mark=at position 0.04 with {\arrow{<}}},
        postaction={decorate}] (0,0) circle (1.2cm);
\draw[green,thick,decoration={markings, mark=at position 0.96 with {\arrow{>}}},
        postaction={decorate}] (0,0) circle (.8cm);
\end{tikzpicture}
\]
Next pinch the region between the defect lines until it splits ($\to$ {\it Basic preparations}).
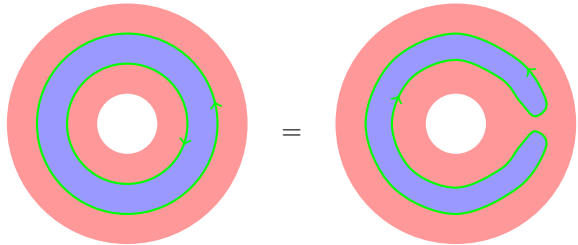
\[
\raisebox{-1.4cm}{\begin{tikzpicture}
\fill[red!40!white] (0,0) circle (1.6cm);
\fill[blue!40!white] (0,0) circle (1.2cm);
\fill[red!40!white] (0,0) circle (0.8cm);
\fill[white] (0,0) circle (0.4cm);
\draw[green,thick,decoration={markings, mark=at position 0.04 with {\arrow{<}}},
        postaction={decorate}] (0,0) circle (1.2cm);
\draw[green,thick,decoration={markings, mark=at position 0.96 with {\arrow{>}}},
        postaction={decorate}] (0,0) circle (.8cm);
\end{tikzpicture}}
\quad
=
\quad
\raisebox{-1.4cm}{\begin{tikzpicture}
\fill[red!40!white] (0,0) circle (1.6cm);
\fill [blue!40!white] plot [smooth cycle, tension=0.6] coordinates {(1.05,0.1) (1.2,0.3) (0.84,0.84) (0,1.2) (-0.84,0.84) (-1.2,0) (-0.84,-0.84) (0,-1.2) (0.84,-0.84) (1.2,-0.3) (1.05,- 0.1) (0.85,-0.3) (0.6,-0.6) (0,-0.85) (-0.6,-0.6) (-0.85,0) (-0.6,0.6) (0,0.85) (0.6,0.6) (0.85,0.3) };
\draw [green, thick ,decoration={markings, mark=at position 0.04 with {\arrow{<}} , mark=at position 0.8 with {\arrow{<}} },
        postaction={decorate} ] plot [smooth cycle, tension=0.6] coordinates {(1.05,0.1) (1.2,0.3) (0.84,0.84) (0,1.2) (-0.84,0.84) (-1.2,0) (-0.84,-0.84) (0,-1.2) (0.84,-0.84) (1.2,-0.3) (1.05,- 0.1) (0.85,-0.3) (0.6,-0.6) (0,-0.85) (-0.6,-0.6) (-0.85,0) (-0.6,0.6) (0,0.85) (0.6,0.6) (0.85,0.3) };
       \fill[white] (0,0) circle (0.4cm);
\end{tikzpicture}
}
\]
Now make the blue bubble disappear\footnote{An US voter may read this as a Republican slogan. I can assure you this was not the intention: I've unsuccessfully tried other options: ``make the red bubble disappear'' could have been more appreciated in traditionally liberal places like US universities but could have been read as anti-communist in Europe;  ``make the black bubble disappear'' would be a great democratic anti-fascist statement in Europe, but it would take a nasty white suprematism flavour in US; ``make the pink bubble disappear'' could at the same time be accused of misogyny and cause a debate on why a given colour should be associated with a gender or a sexual identity. In the end, the best thing is probably not to make any bubble disappear here: are you really so much interested in showing that the centre of {\color{blue}$B$} is isomorphic to the centre of {\color{red}$R$}?}  ($\to$ {\it Basic preparations}):
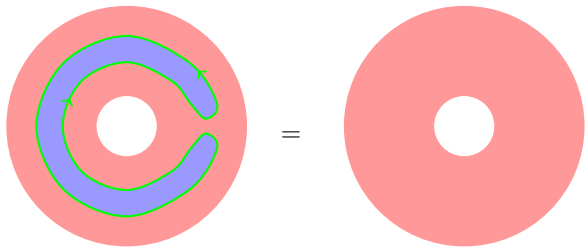
\[
\raisebox{-1.4cm}{
\begin{tikzpicture}
\fill[red!40!white] (0,0) circle (1.6cm);
\fill [blue!40!white] plot [smooth cycle, tension=0.6] coordinates {(1.05,0.1) (1.2,0.3) (0.84,0.84) (0,1.2) (-0.84,0.84) (-1.2,0) (-0.84,-0.84) (0,-1.2) (0.84,-0.84) (1.2,-0.3) (1.05,- 0.1) (0.85,-0.3) (0.6,-0.6) (0,-0.85) (-0.6,-0.6) (-0.85,0) (-0.6,0.6) (0,0.85) (0.6,0.6) (0.85,0.3) };
\draw [green, thick ,decoration={markings, mark=at position 0.04 with {\arrow{<}} , mark=at position 0.8 with {\arrow{<}} },
        postaction={decorate} ] plot [smooth cycle, tension=0.6] coordinates {(1.05,0.1) (1.2,0.3) (0.84,0.84) (0,1.2) (-0.84,0.84) (-1.2,0) (-0.84,-0.84) (0,-1.2) (0.84,-0.84) (1.2,-0.3) (1.05,- 0.1) (0.85,-0.3) (0.6,-0.6) (0,-0.85) (-0.6,-0.6) (-0.85,0) (-0.6,0.6) (0,0.85) (0.6,0.6) (0.85,0.3) };
        \fill[white] (0,0) circle (0.4cm);
\end{tikzpicture}}
\quad=
\quad
\raisebox{-1.4cm}{
\begin{tikzpicture}
\fill[red!40!white] (0,0) circle (1.6cm);
\fill[white] (0,0) circle (0.4cm);
\end{tikzpicture}}
\]
Notice that in preparing the morphisms between the centres, the invertibility of the defect line has not been used: invertibility has only been needed in making the bubble disappear, i.e., in showing that those morphisms were actually isomorphisms. So for noninvertible defect lines one still has distinguished morphisms between the centres, which however will generally not be invertible. This does not mean they will not be interesting: a classical example is the character of a finite dimensional representation $V$ of a finite group $G$ obtained this way by looking at the pair $(V,V^\ast)$ as a noninvertible defect line between $\mathbb{K}$-modules (i.e., $\mathbb{K}$-vector spaces) and $\mathbb{K}[G]$-modules. In case you'd like to serve group characters as an aperitif, here is how to prepare them.
\\ \phantom{i}\\
{\it Ingredients:}
\begin{itemize}
\item[-] A characteristic zero algebraically closed field $\mathbb{K}$;
\item[-] A finite group $G$;
\item[-] the symmetric monoidal $(\infty,2)$-category $\mathcal{C}=\mathrm{Alg_2}$ of finite dimensional $\mathbb{K}$-algebras, bimodules and morphisms of bimodules;
\item[-] ${\color{red}R}=\mathbb{K}[G]$, the semisimple symmetric Frobenius algebra given by the group algebra of $G$ with trace the coefficient of the unit element $1_G$ of $G$;
\item[-] the boundary condition given by $(V,V^\ast)$, where $V$ is a finite dimensional linear representation of $G$, seen as $(\mathbb{K}[G],\mathbb{K})$-bimodule, and $V^\ast$ is the linear dual of $V$, naturally seen as a $(\mathbb{K},\mathbb{K}[G])$-bimodule;
\end{itemize}
\vskip .3 cm
{\it Difficulty:} medium.
\\ \phantom{i}\\
{\it Cooking time:} 6 minutes (1 minute to prepare, 5 minutes to check the preparation is correct).
\\ \phantom{i}\\
{\it Preparation:} 
Mix the ingredients together to obtain
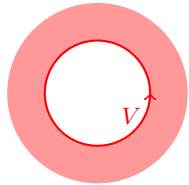
\[
\raisebox{-1cm}{
\begin{tikzpicture}
\fill[red!40!white] (0,0) circle (1.2cm);
\fill[white] (0,0) circle (0.7cm);
\draw[red,thick,decoration={markings, mark=at position 1 with {\arrow{<}}},
        postaction={decorate}] (0,0) circle (.7cm);
        \node[red] at (.45,-.3){\small{$V$}};
\end{tikzpicture}}
\]
The character of $V$ is ready to be served. Just look at the above diagram as a morphism from 
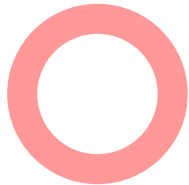
\[
\raisebox{-1cm}{
\begin{tikzpicture}
\fill[red!40!white] (0,0) circle (1.2cm);
\fill[white] (0,0) circle (0.8cm);
\end{tikzpicture}}
\]
to the void to read it as a morphism of $\mathbb{K}$-vector spaces
\[
\mathbb{K}[G]/\bigl[\mathbb{K}[G],\mathbb{K}[G]\bigr]\cong
 \mathbb{K}[G]\otimes^{\circlearrowleft}_{\mathbb{K}[G]}:=
 \mathbb{K}[G]\otimes_{\mathbb{K}[G]\otimes \mathbb{K}[G]^{\mathrm{op}}}\mathbb{K}[G] \to \mathbb{K},
\]
i.e., as a $\mathbb{K}$-valued function on $G$ (as a set) which is constant on conjugacy classes. To see that this function is indeed the character $\chi_V$ of $V$, shake a bit (do not stir!) to obtain
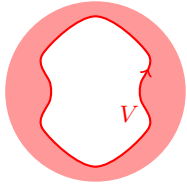
\[
\raisebox{-1cm}{
\begin{tikzpicture}
\fill[red!40!white] (0,0) circle (1.2cm);
        \fill [white] plot [smooth cycle, tension=0.6] coordinates {(0.6,-0.1) (0.6,0.1) (0.7,0.5) (0,1) (-0.7,0.5) (-0.6,0.1)  (-0.6,-0.1) (-0.7,-0.5) (0,-1) (0.7,-0.5) };
 \draw [red, thick ,decoration={markings, mark=at position 0.04 with {\arrow{<}}  },
        postaction={decorate} ] plot [smooth cycle, tension=0.6] coordinates {(0.6,-0.1) (0.6,0.1) (0.7,0.5) (0,1) (-0.7,0.5) (-0.6,0.1)  (-0.6,-0.1) (-0.7,-0.5) (0,-1) (0.7,-0.5)};
                \node[red] at (.45,-.3){\small{$V$}};
 \end{tikzpicture}}
\]
Now slice along the dashed line
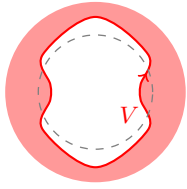
\[
\raisebox{-1cm}{
\begin{tikzpicture}
\fill[red!40!white] (0,0) circle (1.2cm);
        \fill [white] plot [smooth cycle, tension=0.6] coordinates {(0.6,-0.1) (0.6,0.1) (0.7,0.5) (0,1) (-0.7,0.5) (-0.6,0.1)  (-0.6,-0.1) (-0.7,-0.5) (0,-1) (0.7,-0.5) };
 \draw [red, thick ,decoration={markings, mark=at position 0.03 with {\arrow{<}}  },
        postaction={decorate} ] plot [smooth cycle, tension=0.6] coordinates {(0.6,-0.1) (0.6,0.1) (0.7,0.5) (0,1) (-0.7,0.5) (-0.6,0.1)  (-0.6,-0.1) (-0.7,-0.5) (0,-1) (0.7,-0.5)};
                \node[red] at (.45,-.3){\small{$V$}};
                \draw[gray,dashed, line width=0.5pt] (0,0) circle (.76cm);
\end{tikzpicture}}
\]
to identify the original diagram with the composition of
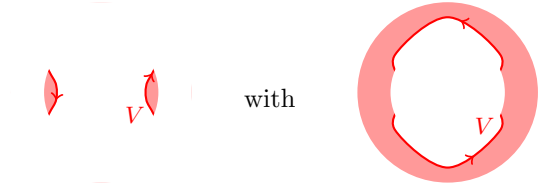
\[
\raisebox{-1cm}{
\begin{tikzpicture}
\fill[red!40!white] (0,0) circle (1.2cm);
        \fill [white] plot [smooth cycle, tension=0.6] coordinates {(0.6,-0.1) (0.6,0.1) (0.7,0.5) (0,1) (-0.7,0.5) (-0.6,0.1)  (-0.6,-0.1) (-0.7,-0.5) (0,-1) (0.7,-0.5) };
 \draw [red, thick ,decoration={markings, mark=at position 0.03 with {\arrow{<}}  , mark=at position 0.5 with {\arrow{<}} },
        postaction={decorate} ] plot [smooth cycle, tension=0.6] coordinates {(0.6,-0.1) (0.6,0.1) (0.7,0.5) (0,1) (-0.7,0.5) (-0.6,0.1)  (-0.6,-0.1) (-0.7,-0.5) (0,-1) (0.7,-0.5)};
                \node[red] at (.45,-.3){\small{$V$}};                
 \path[fill=white,even odd rule]
    (-1.2, -1.2) rectangle (1.2,1.2)
    (0,0)circle[radius=.76];               
\end{tikzpicture}}
\qquad\text{with}\qquad
\raisebox{-1cm}{
\begin{tikzpicture}
\fill[red!40!white] (0,0) circle (1.2cm);
        \fill [white] plot [smooth cycle, tension=0.6] coordinates {(0.6,-0.1) (0.6,0.1) (0.7,0.5) (0,1) (-0.7,0.5) (-0.6,0.1)  (-0.6,-0.1) (-0.7,-0.5) (0,-1) (0.7,-0.5) };
 \draw [red, thick ,decoration={markings, mark=at position 0.2 with {\arrow{<}}   , mark=at position 0.8 with {\arrow{<}} },
        postaction={decorate} ] plot [smooth cycle, tension=0.6] coordinates {(0.6,-0.1) (0.6,0.1) (0.7,0.5) (0,1) (-0.7,0.5) (-0.6,0.1)  (-0.6,-0.1) (-0.7,-0.5) (0,-1) (0.7,-0.5)};
                     \fill[white] (0,0) circle (0.76cm);
                       \node[red] at (.49,-.45){\small{$V$}};
\end{tikzpicture}}
\]
i.e., with the composition
\[
(\mathrm{ev}_V\otimes \mathrm{ev}_V)\circ \tau\circ ({\mathrm{coev}_V}\otimes _{\mathbb{K}[G]}^{\circlearrowleft}\mathrm{coev}_V),
\]
where
\[
\mathrm{ev}_V\colon V^\ast\otimes_{\mathbb{K}[G]} V\to \mathbb{K}
\]
and 
\[
\mathrm{coev}_V\colon \mathbb{K}[G]\to V\otimes V^\ast
\]
are the pairing and the copairing for the representation $V$, respectively, and $\tau$ is the ``rotation'' isomorphism
\begin{align*}
\tau\colon (V\otimes V^\ast)\otimes_{\mathbb{K}[G]}^{\circlearrowleft}
(V\otimes V^\ast)
&\xrightarrow{\sim}
(V^\ast\otimes_{\mathbb{K}[G]}V)\otimes (V^\ast\otimes_{\mathbb{K}[G]} V)\\
(v\otimes \varphi)\otimes_{\mathbb{K}[G]}^{\circlearrowleft}
(w\otimes \psi)&\mapsto (\varphi\otimes_{\mathbb{K}[G]} w)\otimes (\psi\otimes_{\mathbb{K}[G]} v).
\end{align*}
Under the natural isomorphisms $V^\ast\otimes V\cong \mathrm{End}_{\mathbb{K}}(V)\cong V\otimes V^\ast$, the evaluation and the coevaluation of $V$ become the trace
\[
\mathrm{tr}\colon \mathrm{End}_{\mathbb{K}}(V)_{G}^{}\to \mathbb{K}
\]
and the morphism
\[
\rho\colon \mathbb{K}[G]\to \mathrm{End}_{\mathbb{K}}(V)
\]
defining $V$ as a representation of $G$, respectively.
Let $\{e_i\}_{i\in I}$ be a linear basis of $V$ and let $\{e^i\}_{i\in I}$  be the dual basis of $V^\ast$.
Since the canonical isomorphism
\[
\mathbb{K}[G]/\bigl[\mathbb{K}[G],\mathbb{K}[G]\bigr]\cong \mathbb{K}[G]\otimes_{\mathbb{K}[G]}^{\circlearrowleft}
\]
is given by $[x]\mapsto x\otimes 1$, we see that the class function defined by the above construction acts on the element $g$ of $G$ as
\begin{align*}
g&\xmapsto{\sim} g\otimes 1
\xmapsto{(\mathrm{coev}_V\otimes_{\mathbb{K}[G]}^{\circlearrowleft}\mathrm{coev}_V)} \rho(g)\otimes_{\mathbb{K}[G]}^{\circlearrowleft} \rho(1)\\
&=\rho(g)\otimes_{\mathbb{K}[G]}^{\circlearrowleft} \mathrm{id}_V
=(\rho(g)^i_j e_i\otimes e^j)\otimes_{\mathbb{K}[G]}^{\circlearrowleft} (\delta^k_l e_k\otimes e^l)\\
&=\rho(g)^i_j \delta^k_l (e_i\otimes e^j)\otimes_{\mathbb{K}[G]}^{\circlearrowleft}  (e_k\otimes e^l)\xmapsto{\tau}\rho(g)^i_j \delta^k_l  (e^j\otimes_{\mathbb{K}[G]}  e_k)\otimes (e^l\otimes_{\mathbb{K}[G]} e_i)\\
&\xmapsto{\mathrm{ev}_V\otimes \mathrm{ev}_V} \rho(g)^i_j \delta^k_l  \mathrm{tr}(e^j\otimes_{\mathbb{K}[G]}  e_k)\otimes \mathrm{tr}(e^l\otimes_{\mathbb{K}[G]} e_i)\\
&=\rho(g)^i_j \delta^k_l  \delta^j_k\delta^l_i=\rho(g)^i_i=\mathrm{tr}(\rho(g))\\
&=\chi_V(g).
\end{align*}
Another way of checking that we have correctly prepared our character is by making the identification of  
\[
\raisebox{-1cm}{
\begin{tikzpicture}
\fill[red!40!white] (0,0) circle (1.2cm);
\fill[white] (0,0) circle (0.7cm);
\draw[red,thick,decoration={markings, mark=at position 1 with {\arrow{<}}},
        postaction={decorate}] (0,0) circle (.7cm);
        \node[red] at (.45,-.3){\small{$V$}};
\end{tikzpicture}}
\]
with a function on $G$ (constant on conjugacy calsses) manifest. This is done by looking at $\mathbb{K}[G]$ as boundary condition for the red object, by thinking of it as a left $\mathbb{K}$ and a right $\mathbb{K}[G]$-module and then using an element
\[
g\in G\subseteq \mathbb{K}[G]=\mathrm{Hom}_{\mathbb{K},\mathbb{K}[G]}(\mathbb{K}[G],\mathbb{K}[G])
\]
as a defect point on the boundary. The value of our function on the element $g$ is then given by
\[
\raisebox{-1cm}{
\begin{tikzpicture}
\fill[red!40!white] (0,0) circle (1.2cm);
\fill[white] (0,0) circle (0.7cm);
\draw[red,thick,decoration={markings, mark=at position 1 with {\arrow{<}}},
        postaction={decorate}] (0,0) circle (.7cm);
        \node[red] at (.45,-.3){\small{$V$}};
\draw[red,thick,decoration={markings, mark=at position 1 with {\arrow{<}}},
        postaction={decorate}] (0,0) circle (1.2cm);
        \node[red] at (-1.2,0){\textbullet};
        \node[red] at (1.2,1){\footnotesize{$\mathbb{K}[G]$}};
        \node[red] at (-1.45,0){\footnotesize{$g$}};
\end{tikzpicture}}
\]
Now squeeze the annulus (this is a particular instance of pushing a defect line to the boundary ($\to$ {\it Basic preparations}), for a not necessarily invertible defect line):
\[
\raisebox{-1cm}{
\begin{tikzpicture}
\fill[red!40!white] (0,0) circle (1.2cm);
\fill[white] (0,0) circle (0.7cm);
\draw[red,thick,decoration={markings, mark=at position 1 with {\arrow{<}}},
        postaction={decorate}] (0,0) circle (.7cm);
        \node[red] at (.45,-.3){\small{$V$}};
\draw[red,thick,decoration={markings, mark=at position 1 with {\arrow{<}}},
        postaction={decorate}] (0,0) circle (1.2cm);
        \node[red] at (-1.2,0){\textbullet};
        \node[red] at (1.2,1){\footnotesize{$\mathbb{K}[G]$}};
        \node[red] at (-1.55,0){\footnotesize{$\rho(g)$}};
\end{tikzpicture}}=
\raisebox{-1cm}{
\begin{tikzpicture}
        \node[red] at (1.1,-1){\small{$V$}};
\draw[red,thick,decoration={markings, mark=at position 1 with {\arrow{<}}},
        postaction={decorate}] (0,0) circle (1.2cm);
        \node[red] at (-1.2,0){\textbullet};
        \node[red] at (-1.55,0){\footnotesize{$\rho(g)$}};
\end{tikzpicture}}=
\mathrm{tr}(\rho(g))=\chi_V(g).
\]
Here, in the first equality we used that the canonical isomorphism of $(\mathbb{K},\mathbb{K})$-bimodules $\mathbb{K}[G]\otimes_{\mathbb{K}[G]}V\cong V$ identifies the isomorphism $g\otimes \mathrm{id}_V$ of $\mathbb{K}[G]\otimes_{\mathbb{K}[G]}V$ with the isomorphism $\rho(g)$ of $V$, since
\[
(g\otimes \mathrm{id}_V)(1\otimes v)=g\otimes v= 1\otimes \rho(g)(v).
\]
In the second equality we used that
\[
\raisebox{-.5cm}{
\begin{tikzpicture}
\node[red] at (1.8,-.7){\small{$V$}};
\draw[red,thick,decoration={markings, mark=at position .5 with {\arrow{<}}},
        postaction={decorate}] (0,0) arc  [
        start angle=210,
        end angle=330,
        x radius=1.2cm,
        y radius =1.2cm
    ]   ;
\end{tikzpicture}}\,\,\,\colon  \mathbb{K}\to V\otimes V^\ast
\]
and
\[
\raisebox{-.2cm}{
\begin{tikzpicture}
\node[red] at (0.3,.6){\small{$V$}};
\draw[red,thick,decoration={markings, mark=at position .5 with {\arrow{>}}},
        postaction={decorate}] (0,0) arc  [
        start angle=150,
        end angle=30,
        x radius=1.2cm,
        y radius =1.2cm
    ]   ;
\end{tikzpicture}}\,\,\,\colon V\otimes V^\ast=V^{\ast\ast}\otimes V\to \mathbb{K}
\]
are the coevaluation of the $(\mathbb{K},\mathbb{K})$-bimodule $V$ and the evaluation of the $(\mathbb{K},\mathbb{K})$-bimodule $V^\ast$, respectively (where we implicitly used the canonical isomorphism $V^{\ast\ast}\cong V$). This gives
\begin{align*}
\raisebox{-1cm}{
\begin{tikzpicture}
        \node[red] at (1.1,-1){\small{$V$}};
\draw[red,thick,decoration={markings, mark=at position 1 with {\arrow{<}}},
        postaction={decorate}] (0,0) circle (1.2cm);
        \node[red] at (-1.2,0){\textbullet};
        \node[red] at (-1.55,0){\footnotesize{$\rho(g)$}};
\end{tikzpicture}}
&=\mathrm{ev}_{V^\ast}((\rho(g)\otimes \mathrm{id})e_j\otimes e^j)=\mathrm{ev}_{V^\ast}(\rho(g)^i_j e_i\otimes e^j)\\
&=\rho(g)^i_j \delta_i^j=\rho(g)^i_i=\mathrm{tr}(\rho(g)).
\end{align*}
\section{Hors d'oeuvre: nonabelian Fourier transforms}

This is actually a particular case of one of the basic preparations you have learned to cook in full generality. Here we'll see how to serve it so to infuse it with one of its most classic flavours.
\\ \phantom{i}\\
{\it Ingredients:}
\begin{itemize}
\item[-] A characteristic zero algebraically closed field $\mathbb{K}$;
\item[-] A finite group $G$;
\item[-] the symmetric monoidal $(\infty,2)$-category $\mathcal{C}=\mathrm{Alg_2}$ of finite dimensional $\mathbb{K}$-algebras, bimodules and morphisms of bimodules;
\item[-] ${\color{red}R}=\mathbb{K}[G]$, the semisimple symmetric Frobenius algebra given by the group algebra of $G$ with trace the coefficient of the unit element $1_G$ of $G$;
\item[-] ${\color{blue}B}=\mathcal{F}un_{\mathbb{K}}(\hat{G})$, the semisimple symmetric Frobenius algebra of  $\mathbb{K}$-valued functions on the set $\hat{G}$ of isomorphism classes of irreducible representations of $G$ with trace given by the integral with respect to the Plancherel measure $\mu(i)=(\dim_{\mathbb{K}}V_i)^2/|G|$; 
\item[-] the invertible defect line given by $(\bigoplus_{i\in \hat{G}}V_i,\bigoplus_{i\in \hat{G}}V_i^\ast)$, where each representation of $G$ is naturally seen as a left $\mathbb{K}[G]$-module, and the linear dual of a representation is naturally seen as a right $\mathbb{K}[G]$-module; the right and left $\mathcal{F}un_{\mathbb{K}}(\hat{G})$-module structures on $\bigoplus_{i\in \hat{G}}V_i$ and $\bigoplus_{i\in \hat{G}}V_i^\ast$, respectively, are the obvious ones;
\end{itemize}
\vskip .3 cm
{\it Difficulty:} easy.
\\ \phantom{i}\\
{\it Cooking time:} 2 minutes.
\\ \phantom{i}\\
{\it Preparation:} Look at $\mathbb{K}[G]$ as boundary condition for the red object, by thinking of it as a left $\mathbb{K}$ and a right $\mathbb{K}[G]$-module. Pick any 
\[
\alpha\in \mathbb{K}[G]=\mathrm{Hom}_{\mathbb{K},\mathbb{K}[G]}(\mathbb{K}[G],\mathbb{K}[G])
\]
and use it as a defect point on the boundary. Draw
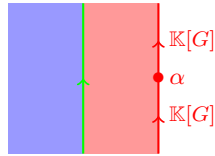
\[
\begin{tikzpicture}
\fill[blue!40!white] (-1,0) rectangle (0,2);
\fill[red!40!white] (0,0) rectangle (1,2);
\draw[green,thick,decoration={markings, mark=at position 0.5 with {\arrow{<}}},
        postaction={decorate}]  (0,0) -- (0,2);
\draw[red,thick,decoration={markings, mark=at position 0.5 with {\arrow{<}}},
        postaction={decorate}]  (1,1) -- (1,2);
\draw[red,thick,decoration={markings, mark=at position 0.5 with {\arrow{<}}},
        postaction={decorate}]   (1,0) -- (1,1);
\node[red] at (1,1){\textbullet};        
\node[red] at (1.45,1.5){\footnotesize{$\mathbb{K}[G]$}};
\node[red] at (1.45,0.5){\footnotesize{$\mathbb{K}[G]$}};
\node[red] at (1.25,1){\small{$\alpha$}};
\end{tikzpicture}
\]

Push the defect line to the boundary ($\to$ {\it Basic preparations}):
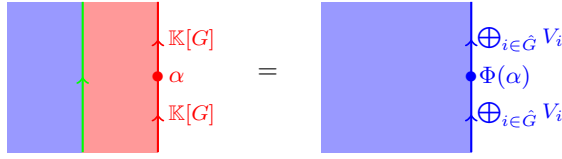
\[
\raisebox{-1cm}{
\begin{tikzpicture}
\fill[blue!40!white] (-1,0) rectangle (0,2);
\fill[red!40!white] (0,0) rectangle (1,2);
\draw[green,thick,decoration={markings, mark=at position 0.5 with {\arrow{<}}},
        postaction={decorate}]  (0,0) -- (0,2);
\draw[red,thick,decoration={markings, mark=at position 0.5 with {\arrow{<}}},
        postaction={decorate}]  (1,1) -- (1,2);
\draw[red,thick,decoration={markings, mark=at position 0.5 with {\arrow{<}}},
        postaction={decorate}]   (1,0) -- (1,1);
\node[red] at (1,1){\textbullet};        
\node[red] at (1.45,1.5){\footnotesize{$\mathbb{K}[G]$}};
\node[red] at (1.45,0.5){\footnotesize{$\mathbb{K}[G]$}};
\node[red] at (1.25,1){\small{$\alpha$}};
\end{tikzpicture}}
\quad =\quad
\raisebox{-1cm}{
\begin{tikzpicture}
\fill[blue!40!white] (-1,0) rectangle (1,2);
\draw[blue,thick,decoration={markings, mark=at position 0.5 with {\arrow{<}}},
        postaction={decorate}]  (1,1) -- (1,2);
\draw[blue,thick,decoration={markings, mark=at position 0.5 with {\arrow{<}}},
        postaction={decorate}]   (1,0) -- (1,1);
\node[blue] at (1,1){\textbullet};        
\node[blue] at (1.65,1.5){\footnotesize{$\bigoplus_{i\in \hat{G}}V_i$}};
\node[blue] at (1.65,0.5){\footnotesize{$\bigoplus_{i\in \hat{G}}V_i$}};
\node[blue] at (1.45,1){\small{$\Phi(\alpha)$}};
\end{tikzpicture}
}
\]
Pick
\[
\Phi(\alpha)\in \mathrm{Hom}_{\mathbb{K},\mathcal{F}un_{\mathbb{K}}(\hat{G})}(\bigoplus_{i\in \hat{G}}V_i,\bigoplus_{i\in \hat{G}}V_i)=\bigoplus_{i\in \hat{G}}\mathrm{End}_{\mathbb{K},\mathbb{K}}(V_i).
\]
You have obtained the isomorphism of algebras (the compatibility with compositions is given by pushing and glueing $\to$ {\it Basic preparations})
\[
\Phi\colon \mathbb{K}[G]\xrightarrow{\sim} \bigoplus_{i\in \hat{G}}\mathrm{End}_{\mathbb{K},\mathbb{K}}(V_i).
\]
known as the nonabelian Fourier transform. Note that the trace on the right is the integral with respect to the Plancherel measure of the traces of the various endomorphisms normalized by the dimension of the representation:
\[
 (\varphi_i)_{i\in \hat{G}}\mapsto \sum_{i\in \hat{G}} \frac{(\dim_{\mathbb{K}}V_i)^2}{|G|} \frac{\mathrm{tr}(\varphi_i)}{\dim_{\mathbb{K}}V_i} =\frac{1}{|G|}{\sum_{i\in \hat{G}} \dim_{\mathbb{K}}V_i\cdot \mathrm{tr}(\varphi_i)}. 
\]
An explicit formula for $\Phi(\alpha)$ is as follows: if $\alpha=\sum_{g\in G}\alpha_g g$, then
\[
(\Phi(\alpha)_i)_{i\in \hat{G}}=\left(\sum_{g\in G}\alpha_g \rho_i(g)\right)_{\!\!\!i\in \hat{G}},
\]
where $\rho_i\colon G\to \mathrm{Aut}_{\mathbb{K},\mathbb{K}}(V_i)$ are the (chosen representatives for the) irreducible representations of $G$.
\\ \phantom{i}\\
{\it Presentation suggestion:} Recall that $\mathbb{K}[G]$ is naturally isomorphic as a $\mathbb{K}$-algebra to the vector space $\mathcal{F}un_{\mathbb{K}}(G)$ of  $\mathbb{K}$-valued functions on the set underlying the group $G$, endowed with the convolution product. The trace of $\mathbb{K}$ is translated by this isomorphism into the evaluation on the unit element $1_G$ of $G$. This way you can present the nonabelian Fourier transform as an algebra isomorphism
\[
\Phi\colon (\mathcal{F}un_{\mathbb{K}}(G),\ast)\xrightarrow{\sim} \bigoplus_{i\in \hat{G}}\mathrm{End}_{\mathbb{K},\mathbb{K}}(V_i)
\]
\\
\subsection{Raw ingredients} The above preparation was so simple since we started with pre-cooked ingredients: we took from the literature shelves the fact that $(\bigoplus_{i\in \hat{G}}V_i,\bigoplus_{i\in \hat{G}}V_i^\ast)$ is an $SO(2)$-homotopy invariant invertible defect line. If we want to start with the raw ingredients, i.e., without knowing this, then what we have is only that $(\bigoplus_{i\in \hat{G}}V_i,\allowbreak\bigoplus_{i\in \hat{G}}V_i^\ast)$ is a defect line between $\mathbb{K}[G]$ and $\mathcal{F}un_{\mathbb{K}}(\hat{G})$, and we have to prove its $SO(2)$-homotopy invariance and its invertibility. The $SO(2)$-homotopy invariance corresponds to the Zorro moves and so, ultimately, to Schur's lemma. As far as concerns the invertibility, verbatim repeating the steps above, we get the algebra homomorphism  $\Phi\colon \mathbb{K}[G]\xrightarrow{\sim} \bigoplus_{i\in \hat{G}}\mathrm{End}_{\mathbb{K},\mathbb{K}}(V_i)$ and proving that our defect line is invertible is \emph{equivalent} to showing this is an isomorphism. This can be done as usual by means of classical characters theory. So, from the point of view of this cookbook, one could say that the main result of character theory is the invertibility of the  $(\bigoplus_{i\in \hat{G}}V_i,\bigoplus_{i\in \hat{G}}V_i^\ast)$ defect.

\subsection{Nonabelian is too spicy? here's the abelian recipe} When $G$ is abelian, the set $\hat{G}$ is the underlying set of the \emph{dual group} of $G$, i.e., of the group of characters 
\[
\chi\colon G\to \mathbb{K}^*
\]
endowed with poinwise multiplication (this corresponds to tensor product of 1-dimensional representations). As all of the irreducible representations of $G$ are 1-dimensional in this case, we have canonical isomorphisms $\mathrm{End}_{\mathbb{K},\mathbb{K}}(V_i)\cong \mathbb{K}$ for any $i\in \hat{G}$ and the Fourier transform becomes an isomorphism
\[
\Phi\colon (\mathcal{F}un_{\mathbb{K}}(G),\ast)\xrightarrow{\sim} (\mathcal{F}un_{\mathbb{K}}(\hat{G}),\cdot).
\]
Written out explicitly, if $\alpha\in \mathcal{F}un_{\mathbb{K}}(G)$, then 
\[
\Phi(\alpha)\colon \hat{G} \to \mathbb{K}
\]
is the function defined by
\[
\Phi(\alpha)\colon \chi\mapsto \sum_{g\in G} \alpha(g)\chi(g).
\]

\subsection{Getting high with higher Fourier transforms}
If you are willing to play with ``foams'' and other techniques from modern haute cuisine, then you need to upgrade from symmetric monoidal $(\infty,2)$-categories to symmetric monoidal $(\infty,3)$-categories. Here you can pick the same kind of ingredients as in the  $(\infty,2)$-categorical case, and prepare the same recipes: they will be beautifully enhanced by the richness of flavour of the $(\infty,3)$-categorical setting. We are not going to dwell into any detail here, but let me at least mention that the one-level-higher version of the pair $(\mathbb{K}[G], \mathcal{F}un_{\mathbb{K}}(\hat{G}))$ is the pair of monoidal categories $({\mathrm{Vect}}_{\mathbb{K}}[G], {\mathrm{Rep}}_{\mathbb{K}}(G))$. The invertible defect surface\footnote{Having moved to $(\infty,3)$-categories, the relevant TQTFs are now 3-dimensional.}  in this case is the pair $(\mathrm{Vect}_{\mathbb{K}}(G\backslash\hskip -3pt\backslash G),\mathrm{Vect}_{\mathbb{K}}(G/\hskip -2pt/G))$ of finite-dimensional $G$-graded $\mathbb{K}$-vector spaces $G$-equivariant for the left (resp., right)  multiplication action of $G$ on itself.

\section{The main course: the Plancherel theorem}
Like with Fourier transforms, here we will start with a very general and verstile recipe, and then we'll cook it with ingredients derived from finite groups to get its most classic flavour.
\\ \phantom{i}\\
{\it Ingredients:} the general ingredients listed at the beginning of the cookbook ($\to$ {\it Ingredients}).
\\ \phantom{i}\\
{\it Difficulty:} easy.
\\ \phantom{i}\\
{\it Cooking time:} 5 minutes.
\\ \phantom{i}\\
{\it Preparation:} Pick $n$ boundary conditions $M_1,\dots, M_n$ for $\color{red}R$ and $n$ defect points $\theta_i:M_i\to M_{i+1}$ on the boundaries, where the index $i$ is taken modulo $n$. Prepare a red disk decorated with these data:
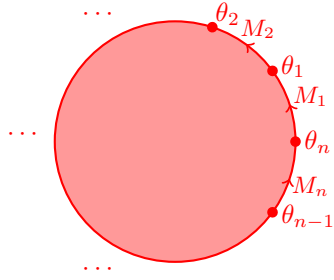
\[
\begin{tikzpicture}
\fill[red!40!white] (0,0) circle (1.6cm);
\draw[red,thick,decoration={markings, mark=at position 0 with {\fill circle [radius=+2pt];},  mark=at position 0.05 with {\arrow{<}} , mark=at position 0.1 with {\fill circle [radius=+2pt];} 
,  mark=at position 0.15 with {\arrow{<}} , mark=at position 0.2 with {\fill circle [radius=+2pt];}
, mark=at position 0.9 with {\fill circle [radius=+2pt];} ,  mark=at position 0.95 with {\arrow{<}} },
        postaction={decorate}] (0,0) circle (1.6cm);
\node[red] at (-1,1.7){$\cdots$};   
\node[red] at (-2,0.1){$\cdots$};     
\node[red] at (-1,-1.7){$\cdots$}; 
\node[red] at (.7,1.7){$\theta_2$};  
\node[red] at (1.57,1){$\theta_1$};   
\node[red] at (1.9,0){$\theta_n$};     
\node[red] at (1.77,-1){$\theta_{n-1}$}; 
\node[red] at (1.8,.6){\small{$M_1$}}; 
\node[red] at (1.8,-.6){\small{$M_n$}};
\node[red] at (1.1,1.5){\small{$M_2$}};
\end{tikzpicture}
\]
Add a blue bubble in the middle ($\to$ {\it Basic preparations}):
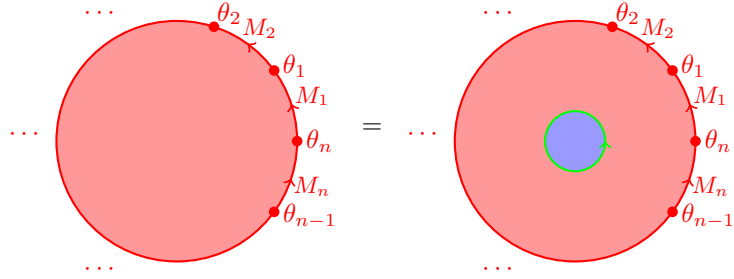
\[
\raisebox{-2cm}{
\begin{tikzpicture}
\fill[red!40!white] (0,0) circle (1.6cm);
\draw[red,thick,decoration={markings, mark=at position 0 with {\fill circle [radius=+2pt];},  mark=at position 0.05 with {\arrow{<}} , mark=at position 0.1 with {\fill circle [radius=+2pt];} 
,  mark=at position 0.15 with {\arrow{<}} , mark=at position 0.2 with {\fill circle [radius=+2pt];}
, mark=at position 0.9 with {\fill circle [radius=+2pt];} ,  mark=at position 0.95 with {\arrow{<}} },
        postaction={decorate}] (0,0) circle (1.6cm);
\node[red] at (-1,1.7){$\cdots$};   
\node[red] at (-2,0.1){$\cdots$};     
\node[red] at (-1,-1.7){$\cdots$}; 
\node[red] at (.7,1.7){$\theta_2$};  
\node[red] at (1.57,1){$\theta_1$};   
\node[red] at (1.9,0){$\theta_n$};     
\node[red] at (1.77,-1){$\theta_{n-1}$}; 
\node[red] at (1.8,.6){\small{$M_1$}}; 
\node[red] at (1.8,-.6){\small{$M_n$}};
\node[red] at (1.1,1.5){\small{$M_2$}};
\end{tikzpicture}
}
=
\raisebox{-2cm}{
\begin{tikzpicture}
\fill[red!40!white] (0,0) circle (1.6cm);
\draw[red,thick,decoration={markings, mark=at position 0 with {\fill circle [radius=+2pt];},  mark=at position 0.05 with {\arrow{<}} , mark=at position 0.1 with {\fill circle [radius=+2pt];} 
,  mark=at position 0.15 with {\arrow{<}} , mark=at position 0.2 with {\fill circle [radius=+2pt];}
, mark=at position 0.9 with {\fill circle [radius=+2pt];} ,  mark=at position 0.95 with {\arrow{<}} },
        postaction={decorate}] (0,0) circle (1.6cm);
\node[red] at (-1,1.7){$\cdots$};   
\node[red] at (-2,0.1){$\cdots$};     
\node[red] at (-1,-1.7){$\cdots$}; 
\node[red] at (.7,1.7){$\theta_2$};  
\node[red] at (1.57,1){$\theta_1$};   
\node[red] at (1.9,0){$\theta_n$};     
\node[red] at (1.77,-1){$\theta_{n-1}$}; 
\node[red] at (1.8,.6){\small{$M_1$}}; 
\node[red] at (1.8,-.6){\small{$M_n$}};
\node[red] at (1.1,1.5){\small{$M_2$}};
\fill[blue!40!white] (0,0) circle (.4cm);
\draw[green,thick,decoration={markings, mark=at position 1 with {\arrow{<}}},
        postaction={decorate}] (0,0) circle (.4cm);
\end{tikzpicture}
}
\]
Push the defect line to the boundary ($\to$ {\it Basic preparations})
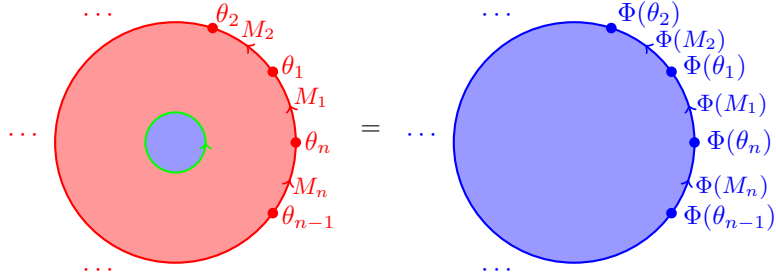
\[
\raisebox{-2cm}{
\begin{tikzpicture}
\fill[red!40!white] (0,0) circle (1.6cm);
\draw[red,thick,decoration={markings, mark=at position 0 with {\fill circle [radius=+2pt];},  mark=at position 0.05 with {\arrow{<}} , mark=at position 0.1 with {\fill circle [radius=+2pt];} 
,  mark=at position 0.15 with {\arrow{<}} , mark=at position 0.2 with {\fill circle [radius=+2pt];}
, mark=at position 0.9 with {\fill circle [radius=+2pt];} ,  mark=at position 0.95 with {\arrow{<}} },
        postaction={decorate}] (0,0) circle (1.6cm);
\node[red] at (-1,1.7){$\cdots$};   
\node[red] at (-2,0.1){$\cdots$};     
\node[red] at (-1,-1.7){$\cdots$}; 
\node[red] at (.7,1.7){$\theta_2$};  
\node[red] at (1.57,1){$\theta_1$};   
\node[red] at (1.9,0){$\theta_n$};     
\node[red] at (1.77,-1){$\theta_{n-1}$}; 
\node[red] at (1.8,.6){\small{$M_1$}}; 
\node[red] at (1.8,-.6){\small{$M_n$}};
\node[red] at (1.1,1.5){\small{$M_2$}};
\fill[blue!40!white] (0,0) circle (.4cm);
\draw[green,thick,decoration={markings, mark=at position 1 with {\arrow{<}}},
        postaction={decorate}] (0,0) circle (.4cm);
\end{tikzpicture}
}
=
\raisebox{-2cm}{
\begin{tikzpicture}
\fill[blue!40!white] (0,0) circle (1.6cm);
\draw[blue,thick,decoration={markings, mark=at position 0 with {\fill circle [radius=+2pt];},  mark=at position 0.05 with {\arrow{<}} , mark=at position 0.1 with {\fill circle [radius=+2pt];} 
,  mark=at position 0.15 with {\arrow{<}} , mark=at position 0.2 with {\fill circle [radius=+2pt];}
, mark=at position 0.9 with {\fill circle [radius=+2pt];} ,  mark=at position 0.95 with {\arrow{<}} },
        postaction={decorate}] (0,0) circle (1.6cm);
\node[blue] at (-1,1.7){$\cdots$};   
\node[blue] at (-2,0.1){$\cdots$};     
\node[blue] at (-1,-1.7){$\cdots$}; 
\node[blue] at (1,1.7){$\Phi(\theta_2)$};  
\node[blue] at (1.87,1){$\Phi(\theta_1)$};   
\node[blue] at (2.2,0){$\Phi(\theta_n)$};     
\node[blue] at (2.07,-1){$\Phi(\theta_{n-1})$}; 
\node[blue] at (2.1,.5){\small{$\Phi(M_1)$}}; 
\node[blue] at (2.1,-.6){\small{$\Phi(M_n)$}};
\node[blue] at (1.55,1.35){\small{$\Phi(M_2)$}};
\end{tikzpicture}
}
\]
That's it. As this may leave too strong an abstract aftertaste, here is the classical recipe for a finite group $G$.
\\ \phantom{i}\\
{\it Ingredients:} same as for the nonabelian Fourier transform ($\to$ {\it Hors d'oeuvre})
\\ \phantom{i}\\
{\it Difficulty:} easy.
\\ \phantom{i}\\
{\it Cooking time:} 3 minutes.
\\ \phantom{i}\\
{\it Preparation:} Look at $\mathbb{K}[G]$ as boundary condition for the red object, by thinking of it as a left $\mathbb{K}$ and a right $\mathbb{K}[G]$-module. Use the the canonical isomorphism $\mathbb{K}[G]\cong (\mathcal{F}un_{\mathbb{K}}(G),\ast)$ to think of $\mathcal{F}un_{\mathbb{K}}(G)$ as a boundary condition for the red object. Chose $M_i=\mathcal{F}un_{\mathbb{K}}(G)$ for any $i=1,\dots,n$. Pick arbitrary functions $\theta_1,\dots,\theta_n\in \mathcal{F}un_{\mathbb{K}}(G)$ and use them as defect points. The general recipe then specialises to
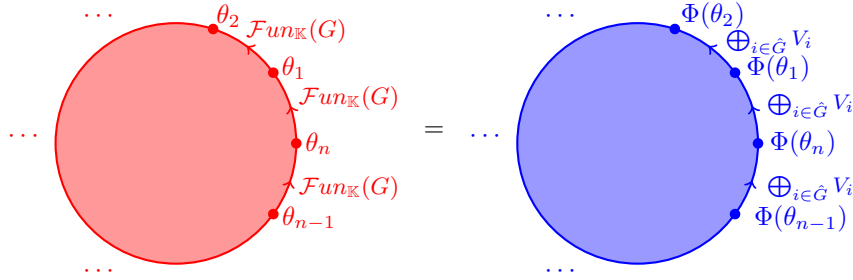
\[
\raisebox{-2cm}{
\begin{tikzpicture}
\fill[red!40!white] (0,0) circle (1.6cm);
\draw[red,thick,decoration={markings, mark=at position 0 with {\fill circle [radius=+2pt];},  mark=at position 0.05 with {\arrow{<}} , mark=at position 0.1 with {\fill circle [radius=+2pt];} 
,  mark=at position 0.15 with {\arrow{<}} , mark=at position 0.2 with {\fill circle [radius=+2pt];}
, mark=at position 0.9 with {\fill circle [radius=+2pt];} ,  mark=at position 0.95 with {\arrow{<}} },
        postaction={decorate}] (0,0) circle (1.6cm);
\node[red] at (-1,1.7){$\cdots$};   
\node[red] at (-2,0.1){$\cdots$};     
\node[red] at (-1,-1.7){$\cdots$}; 
\node[red] at (.7,1.7){$\theta_2$};  
\node[red] at (1.57,1){$\theta_1$};   
\node[red] at (1.9,0){$\theta_n$};     
\node[red] at (1.77,-1){$\theta_{n-1}$}; 
\node[red] at (2.3,.6){\small{$\mathcal{F}un_{\mathbb{K}}(G)$}}; 
\node[red] at (2.3,-.6){\small{$\mathcal{F}un_{\mathbb{K}}(G)$}};
\node[red] at (1.6,1.5){\small{$\mathcal{F}un_{\mathbb{K}}(G)$}};
\end{tikzpicture}
}
=
\raisebox{-2cm}{
\begin{tikzpicture}
\fill[blue!40!white] (0,0) circle (1.6cm);
\draw[blue,thick,decoration={markings, mark=at position 0 with {\fill circle [radius=+2pt];},  mark=at position 0.05 with {\arrow{<}} , mark=at position 0.1 with {\fill circle [radius=+2pt];} 
,  mark=at position 0.15 with {\arrow{<}} , mark=at position 0.2 with {\fill circle [radius=+2pt];}
, mark=at position 0.9 with {\fill circle [radius=+2pt];} ,  mark=at position 0.95 with {\arrow{<}} },
        postaction={decorate}] (0,0) circle (1.6cm);
\node[blue] at (-1,1.7){$\cdots$};   
\node[blue] at (-2,0.1){$\cdots$};     
\node[blue] at (-1,-1.7){$\cdots$}; 
\node[blue] at (1,1.7){$\Phi(\theta_2)$};  
\node[blue] at (1.87,1){$\Phi(\theta_1)$};   
\node[blue] at (2.2,0){$\Phi(\theta_n)$};     
\node[blue] at (2.17,-1){$\Phi(\theta_{n-1})$}; 
\node[blue] at (2.3,.5){\footnotesize{$\bigoplus_{i\in \hat{G}}V_i$}}; 
\node[blue] at (2.3,-.6){\footnotesize{$\bigoplus_{i\in \hat{G}}V_i$}};
\node[blue] at (1.75,1.35){\footnotesize{$\bigoplus_{i\in \hat{G}}V_i$}};
\end{tikzpicture}
}
\]
i.e., to the identity
\[
(\theta_1*\theta_2*\cdots *\theta_n)(1_G)=\frac{1}{|G|}\sum_{i\in \hat{G}} (\dim V_i)\mathrm{tr}(\Phi(\theta_1)_i\circ \Phi(\theta_2)_i\circ\cdots\circ  \Phi(\theta_n)_i).
\]
The convolution product is explicitly given by
\[
(\theta_1*\theta_2)(h)=\sum_{g\in G} \theta_1(g)\theta_2(g^{-1}h)
\]
and it is associative. So, inductively one has
\[
(\theta_1*\theta_2*\cdots *\theta_n)(h)=\sum_{g_1,\dots, g_{n-1}\in G} \theta_1(g_1) \theta_2(g_2)\cdots \theta_{n-1}(g_{n-1})\theta_n(g_{n-1}^{-1}\cdots g_2^{-1}g_1^{-1}h).
\]
Hence one gets the identity
\[
\sum_{g_1,\dots, g_{n-1}\in G} \!\!\!\!\!\theta_1(g_1) \cdots \theta_{n-1}(g_{n-1})\theta_n(g_{n-1}^{-1}\cdots g_1^{-1})=\frac{1}{|G|}\sum_{i\in \hat{G}} (\dim V_i)\mathrm{tr}(\Phi(\theta_1)_i\circ \cdots\circ  \Phi(\theta_n)_i)
\]
giving the $n$-point Plancherel theorem. The traditional recipe for this has $n=2$, and therefore reads
\[
\sum_{g\in G} \theta_1(g)\theta_2(g^{-1})=\frac{1}{|G|}\sum_{i\in \hat{G}} (\dim V_i)\mathrm{tr}(\Phi(\theta_1)_i\circ \Phi(\theta_2)_i)
\]
\\ \phantom{i}\\
{\it Presentation suggestion:} For any fixed $k_0$, by setting $k_i=k_{i-1}g_i$ for $i=1,\dots,n$, you have the identity
\begin{align*}
\sum_{g_1,\dots, g_{n-1}\in G} \!\!\!\!\!\theta_1(g_1) \cdots & \theta_{n-1}(g_{n-1})\theta_n(g_{n-1}^{-1}\cdots g_1^{-1})\\
&= \sum_{k_1,\dots, k_{n-1}\in G} \theta_1(k_0^{-1}k_1) \theta_2(k_1^{-1}k_2)\cdots \theta_{n-1}(k_{n-2}^{-1}k_{n-1})\theta_n(k_{n-1}^{-1}k_0).
\end{align*}
Summing over $k_0$ you therefore get
\begin{align*}
\sum_{g_1,\dots, g_{n-1}\in G} \!\!\!\!\!\theta_1(g_1) \cdots & \theta_{n-1}(g_{n-1})\theta_n(g_{n-1}^{-1}\cdots g_1^{-1})\\
&=\frac{1}{|G|} \sum_{k_0,k_1,\dots, k_{n-1}\in G} \theta_1(k_0^{-1}k_1) \theta_2(k_1^{-1}k_2)\cdots \theta_{n-1}(k_{n-2}^{-1}k_{n-1})\theta_n(k_{n-1}^{-1}k_0)
\end{align*}
and, renaming $k_i=g_{i+1}$, the $n$-point Plancherel theorem takes the elegant cyclic form
\[
\sum_{g_1,\dots, g_{n}\in G} \theta_1(g_1^{-1}g_2) \cdots \theta_{n-1}(g_{n-1}^{-1}g_{n})\theta_n(g_{n}^{-1}g_1)
=
\sum_{i\in \hat{G}} (\dim V_i)\mathrm{tr}(\Phi(\theta_1)_i\circ \cdots\circ  \Phi(\theta_n)_i)
\]
\subsection{The inverse Fourier transform} Writing an explicit expression for
\[
\Phi^{-1}\colon \bigoplus_{i\in \hat{G}}\mathrm{End}_{\mathbb{K},\mathbb{K}}(V_i)\xrightarrow{\sim}\mathbb{K}[G]
\]
is slightly less immediate than one may expect. Namely, repeating the steps to prepare the Fourier transform, with $\color{red}R$ and $\color{blue}B$ exchanged and starting with $\bigoplus_{i\in \hat{G}}V_i$ on the boundary, one does not end up with an explicit $\mathbb{K}[G]$ on the boundary as one would have hoped for, but rather with $\bigoplus_{i\in \hat{G}}\mathrm{End}_{\mathbb{K},\mathbb{K}}(V_i)$ on the boundary. And the identification of this with $\mathbb{K}[G]$ is precisely the seeked for $\Phi^{-1}$, which therefore remains unexplicited if we do so. So here's the chef's trick. Knowing an element $\alpha\in \mathbb{K}[G]$ means knowing its coefficients $\alpha_g$, and the coefficient of $g$ in $\alpha$ is nothing but the coefficient of $1_G$ in $g^{-1}\cdot \alpha$. Therefore we find
\[
\hskip -.5cm
(\Phi^{-1}((\varphi_i)_{i\in \hat{G}})_g= 
\raisebox{-2cm}{
\begin{tikzpicture}
\fill[red!40!white] (0,0) circle (1.6cm);
\draw[red,thick,decoration={markings, mark=at position 0 with {\fill circle [radius=+2pt];},  mark=at position 0.12 with {\arrow{<}} , mark=at position 0.2 with {\fill circle [radius=+2pt];} 
,  mark=at position 0.55 with {\arrow{<}} 
},
        postaction={decorate}] (0,0) circle (1.6cm);
\node[red] at (-1,1.65){\footnotesize{$\mathbb{K}[G]$}};   
\node[red] at (.95,1.7){$g^{-1}$};   
\node[red] at (2.77,0){$\Phi^{-1}((\varphi_i)_{i\in \hat{G}})$};     
\node[red] at (1.8,.9){\footnotesize{$\mathbb{K}[G]$}}; 
\end{tikzpicture}
}
=
\raisebox{-2cm}{
\begin{tikzpicture}
\fill[blue!40!white] (0,0) circle (1.6cm);
\draw[blue,thick,decoration={markings, mark=at position 0 with {\fill circle [radius=+2pt];},  mark=at position 0.12 with {\arrow{<}} , mark=at position 0.2 with {\fill circle [radius=+2pt];} 
,  mark=at position 0.55 with {\arrow{<}} 
},
        postaction={decorate}] (0,0) circle (1.6cm);
\node[blue] at (-1.1,1.65){\footnotesize{$\bigoplus_{i\in \hat{G}}V_i$}};   
\node[blue] at (1.05,1.7){$\Phi(g^{-1})$};   
\node[blue] at (2.27,0){$(\varphi_i)_{i\in \hat{G}}$};     
\node[blue] at (2.1,.9){\footnotesize{$\bigoplus_{i\in \hat{G}}V_i$}}; 
\end{tikzpicture}
}
\]
From this, recalling that
\[
(\Phi(g^{-1})_i)_{i\in \hat{G}}=\left(\rho_i(g^{-1})\right)_{i\in \hat{G}},
\]
we get the explicit formula
\[
(\Phi^{-1}((\varphi_i)_{i\in \hat{G}})_g=
\frac{1}{|G|}\sum_{i\in \hat{G}} (\dim_{\mathbb{K}}V_i)\mathrm{tr}\left(\rho_i(g^{-1})\circ \varphi_i\right).
\]
In the abelian case the inverse transform is immediately seen to reduce to the following. If $f\in \mathcal{F}un_{\mathbb{K}}(\hat{G})$ then 
\[
\Phi^{-1}(f)\colon G \to \mathbb{K}
\]
is the function defined by
\[
\Phi^{-1}(f)\colon g\mapsto 
\frac{1}{|G|}\sum_{\chi\in \hat{G}} \chi(g^{-1}) f(\chi).
\]

\section{Sweet ending: the 1d Ising model}

This is actually a very particular case of the $n$-point Plancherel theorem. But if you serve the classic $n=2$ Plancherel theorem as main, then this specialisation of the arbitrary $n$ case can be a very nice ending (especially if you have statistical mechanics scholars over for dinner), as it realises the 1-dimensional Ising model as a boundary theory for a fully extended 2-dimensional TQFT. The analogous statement is true in any dimension, but already the 2-dimensional Ising model realisation as a boundary theory for a fully extended 3d TQFT goes far beyond the cooking abilities assumed here. Should you be interested in it, I'm providing references in the list of suggested readings at the end of this note.
\\ \phantom{i}\\
{\it Ingredients:}
\begin{itemize}
\item[-] The field $\mathbb{C}$ of complex numbers;
\item[-] A complex number $\beta$;
\item[-] The finite group $\bm{\mu}_2=\{1,-1\}\subseteq \mathbb{C}^\ast$ (the multiplicative group of square roots of 1 in $\mathbb{C}$);
\end{itemize}
\vskip .3 cm
{\it Difficulty:} easy.
\\ \phantom{i}\\
{\it Cooking time:} 2 minutes.
\\ \phantom{i}\\
{\it Preparation:} 
To prepare the 1d Ising model as a boundary 2d TQFT, notice that $\bm{\mu}_2$ acts on $\mathbb{C}$ by multiplication and define $\theta_\beta\in \mathcal{F}un(\bm{\mu}_2,\mathbb{C})$ as
\[
\theta_\beta(\sigma)=e^{\beta\sigma},
\]
i.e., $\theta_\beta(\pm1)=e^{\pm\beta}$. Using $\sigma^{-1}=\sigma$ for any $\sigma\in \bm{\mu}_2$, compute
\begin{align*}
\sum_{\sigma_1,\sigma_2,\dots, \sigma_{n}\in \bm{\mu}_2} \theta_\beta(\sigma_1^{-1}\sigma_2) \cdots \theta_\beta(\sigma_{n}^{-1}\sigma_1)&=
\sum_{\sigma_1,\sigma_2,\dots, \sigma_{n}\in \{1,-1\}} \theta_\beta(\sigma_1\sigma_2) \cdots \theta_\beta(\sigma_{n}\sigma_1)\\
&=\sum_{\sigma_1,\sigma_2,\dots, \sigma_{n}\in \{1,-1\}}e^{\beta\sigma_1\sigma_2} \cdots e^{\beta\sigma_{n}\sigma_1}\\
&=\sum_{\sigma_1,\sigma_2,\dots, \sigma_{n}\in \{1,-1\}}e^{\beta\sum_{j=1}^n\sigma_j\sigma_{j+1}}, 
\end{align*}
where in the sum over $j$ the indices are taken modulo $n$. In the rightmost term you have obtained \emph{partition function} $Z_\beta(n)$ of the periodic 1d Ising model with $n$ nodes. Now recall that the two irreducible complex representations of $\bm{\mu}_2$ are the trivial representation and the defining (or sign) representation, and compute
\[
\Phi(\theta_\beta)=\left(
\begin{matrix}
e^\beta+e^{-\beta} & 0\\
0 & e^\beta-e^{-\beta}
\end{matrix}
\right)
=
\left(
\begin{matrix}
2\,\mathrm{cosh}\, \beta & 0\\
0 & 2\,\mathrm{sinh}\, \beta
\end{matrix}
\right).
\]
Complete the preparation by writing
\[
Z_\beta(n)=
\raisebox{-2cm}{
\begin{tikzpicture}
\fill[red!40!white] (0,0) circle (1.6cm);
\draw[red,thick,decoration={markings, mark=at position 0 with {\fill circle [radius=+2pt];},  mark=at position 0.05 with {\arrow{<}} , mark=at position 0.1 with {\fill circle [radius=+2pt];} 
,  mark=at position 0.15 with {\arrow{<}} , mark=at position 0.2 with {\fill circle [radius=+2pt];}
, mark=at position 0.9 with {\fill circle [radius=+2pt];} ,  mark=at position 0.95 with {\arrow{<}} },
        postaction={decorate}] (0,0) circle (1.6cm);
\node[red] at (-1,1.7){$\cdots$};   
\node[red] at (-2,0.1){$\cdots$};     
\node[red] at (-1,-1.7){$\cdots$}; 
\node[red] at (.7,1.7){$\theta_\beta$};  
\node[red] at (1.57,1){$\theta_\beta$};   
\node[red] at (1.9,0){$\theta_\beta$};     
\node[red] at (1.77,-1){$\theta_{n-1}$}; 
\node[red] at (2.3,.6){\small{$\mathcal{F}un_{\mathbb{C}}(\bm{\mu}_2)$}}; 
\node[red] at (2.3,-.6){\small{$\mathcal{F}un_{\mathbb{C}}(\bm{\mu}_2)$}};
\node[red] at (1.6,1.5){\small{$\mathcal{F}un_{\mathbb{C}}(\bm{\mu}_2)$}};
\end{tikzpicture}
}
=
\raisebox{-2cm}{
\begin{tikzpicture}
\fill[blue!40!white] (0,0) circle (1.6cm);
\draw[blue,thick,decoration={markings, mark=at position 0 with {\fill circle [radius=+2pt];},  mark=at position 0.05 with {\arrow{<}} , mark=at position 0.1 with {\fill circle [radius=+2pt];} 
,  mark=at position 0.15 with {\arrow{<}} , mark=at position 0.2 with {\fill circle [radius=+2pt];}
, mark=at position 0.9 with {\fill circle [radius=+2pt];} ,  mark=at position 0.95 with {\arrow{<}} },
        postaction={decorate}] (0,0) circle (1.6cm);
\node[blue] at (-1,1.7){$\cdots$};   
\node[blue] at (-2,0.1){$\cdots$};     
\node[blue] at (-1,-1.7){$\cdots$}; 
\node[blue] at (1,1.9){\tiny{$\left(
\begin{smallmatrix}
2\mathrm{cosh}\, \beta & 0\\
0 & 2\mathrm{sinh}\, \beta
\end{smallmatrix}
\right)$}};  
\node[blue] at (2.4,1){\tiny{$\left(
\begin{smallmatrix}
2\mathrm{cosh}\, \beta & 0\\
0 & 2\mathrm{sinh}\, \beta
\end{smallmatrix}
\right)$}};   
\node[blue] at (2.7,0){\tiny{$\left(
\begin{smallmatrix}
2\mathrm{cosh}\, \beta & 0\\
0 & 2\mathrm{sinh}\, \beta
\end{smallmatrix}
\right)$}};     
\node[blue] at (2.4,-1){\tiny{$\left(
\begin{smallmatrix}
2\mathrm{cosh}\, \beta & 0\\
0 & 2\mathrm{sinh}\, \beta
\end{smallmatrix}
\right)$}}; 
\node[blue] at (2.5,.5){\footnotesize{$\mathbb{C}_{\mathrm{triv}}\oplus \mathbb{C}_{\mathrm{sgn}}$}}; 
\node[blue] at (2.5,-.6){\footnotesize{$\mathbb{C}_{\mathrm{triv}}\oplus \mathbb{C}_{\mathrm{sgn}}$}};
\node[blue] at (1.95,1.35){\footnotesize{$\mathbb{C}_{\mathrm{triv}}\oplus \mathbb{C}_{\mathrm{sgn}}$}};
\end{tikzpicture}
}
\]
i.e.,
\[
Z_\beta(n)
 =
(2\,\mathrm{cosh}\, \beta)^n+(2\,\mathrm{sinh}\, \beta)^n.
\]

\section{Acknowledgments and further readings}
This little cookbook could not exist without several conversations on fully extended TQFTs I had over the years with Nils Carqueville and Alessandro Valentino. It is no exaggeration to say that everything I know about fully extended TQFTs is thanks to them, so it is a pleasure to end this note by thanking them for all the beautiful mathematics they taught to me (among many other things too personal to be mentioned here). It is also a pleasure to thank the readers of the first version of this cookbook for their enthusiastic feedback, for having spotted a few inaccuracies and too many italianisms, and for having inspired me to add a (little) more advanced recipe in this revised version.  
\vskip .4 cm 
If you enjoyed the recipes presented here and are curious of preparing yourself more sophisticated ones, there are basically two ways. The first, and more effective, is to pester Alessandro and Nils as I did. If you feel too shy for this, then I can recommend a few pleasant text to read. The list here below is necessarily incomplete, I apologise for any omission.

\renewcommand{\refname}{}

\addtocontents{toc}{\protect\setcounter{tocdepth}{-1}}


\nocite{*} 
\begin{thebibliography}{99} 
   
\bibitem{atiyah} Michael  Atiyah, {\it Topological  quantum  field  theories},  Inst.  Hautes \'Etudes  Sci.  Publ.Math. (1988), no. 68, 175--186. 

\bibitem{bakalov-kirillov} Bojko Bakalov and Alexander Kirillov, Jr. {\it Lectures on Tensor Categories and Modular Functors}, University Lecture Series
Volume: 21; American Mathematical Society, 2001; 221 pp. 

\bibitem{baez-dolan} John C. Baez and James Dolan,{\it Higher-dimensional algebra and topological quantum field theory}, J. Math. Phys.36 (1995), no. 11, 6073--6105.

\bibitem{bartlett-et-al} Bruce  Bartlett,  Christopher  L.  Douglas,  Christopher  J.  Schommer-Pries,  and  Jamie Vicary, {\it Modular categories as representations of the 3-dimensional bordism 2-category}, {\tt  arXiv:1509.06811}

\bibitem{campbell-ponto} Jonathan A. Campbell and Kate Ponto, {\it Topological Hochschild Homology and Higher Characteristics}, Algebr. Geom. Topol. 19 (2019) 965--1017.

\bibitem{carqueville-defect} Nils Carqueville, {\it Lecture notes on 2-dimensional defect TQFT},  Banach Center Publications 114 (2018), 49--84. 

\bibitem{carqueville-runkel} Nils Carqueville and Ingo Runkel, {\it Introductory lectures on topological quantum field theory},  Banach Center Publications 114 (2018), 9--47.

\bibitem{carqueville-runkel-schaumann} Nils Carqueville, Ingo Runkel and Gregor Schaumann {\it Line and surface defects in Reshetikhin-Turaev TQFT}, {\tt  arXiv:1710.10214}

 
\bibitem{douglas-et-al} Christopher L. Douglas, Christopher J. Schommer-Pries, and Noah Snyder, {\it Dualizable tensor categories}, {\tt arXiv 1312.7188}

\bibitem{freed} Daniel  S.  Freed, {\it The  cobordism  hypothesis},  Bull.  Amer.  Math.  Soc.  (N.S.)50(2013),  no.  1,  57--92.

\bibitem{fhlt} Daniel S. Freed,  Michael J. Hopkins,  Jacob Lurie,  and Constantin Teleman, {\it Topological  quantum  field theories  from  compact  Lie  groups}, A celebration of the mathematical legacy of Raoul Bott, CRM Proc.Lecture Notes, vol. 50, Amer. Math. Soc., Providence, RI, 2010, pp. 367--403.

\bibitem{freed-quinn} Daniel S. Freed and Frank Quinn, {\it Chern-Simons theory with finite gauge group}, Comm. Mth. Phys.156 (1993), no. 3, 435--472.

\bibitem{freed-teleman1} Daniel S. Freed and Constantin Teleman, {\it Relative quantum field theory}, Comm. Math. Phys. 326 (2014), no. 2, 459--476. 

\bibitem{freed-teleman2} Daniel S. Freed and Constantin Teleman, {\it Topological dualities in the Ising model}, {\tt arXiv:1806.00008}

\bibitem{fiorenza-valentino} Domenico Fiorenza and Alessandro Valentino, {\it Boundary Conditions for Topological Quantum Field Theories, Anomalies and Projective Modular Functors}. Commun. Math. Phys. 338, 1043--1074 (2015).

\bibitem{gwilliam-scheimbauer} Owen Gwilliam and Claudia Scheimbauer, {\it Duals and adjoints in higher Morita categories}, {\tt arXiv:1804.10924}

\bibitem{khovanov} Mikhail Khovanov, {\it Categorifications from planar diagrammatics.} Jpn. J. Math. 5, 153–181 (2010). 

\bibitem{lurie} Jacob Lurie, {\it On  the  classification  of  topological  field  theories},  Current  developments  in mathematics, 2008, Int. Press, Somerville, MA, 2009, pp. 129--280.  

\bibitem{moore-segal} Gregory W. Moore and Graeme Segal, {\it D-branes and K-theory in 2D topological field theory}, {\tt arXiv:hep-th/0609042}


\bibitem{quinn} Frank Quinn, {\it Lectures on axiomatic topological quantum field theory}, Geometry and quantum field theory(Park  City,  UT,  1991),  IAS/Park  City  Math.  Ser.,  vol.  1,  Amer.  Math.  Soc.,  Providence,  RI,  1995, pp. 323--453.

\bibitem{schommer-pries} Christopher  John  Schommer-Pries, {\it The  classification  of  two-dimensional  extended topological  field  theories},  ProQuest  LLC,  Ann  Arbor,  MI,  2009,  Thesis  (Ph.D.) -- University of California, Berkeley.  

\bibitem{severa} Pavol Severa, {\it (Non-)Abelian  Kramers-Wannier  Duality  And  Topological  Field  Theory}, Journal of HighEnergy Physics 2002 (2002), no. 05, 049--049.

\bibitem{segal} Graeme Segal, {\it The  definition  of  conformal  field  theory},  Topology,  geometry and quantum field theory, London Math. Soc. Lecture Note Ser., vol. 308, Cambridge Univ. Press, Cambridge, 2004, pp. 421--577.

\bibitem{teleman}Constantin  Teleman, {\it Five  lectures  on  topological  field  theory},  Geometry  and  Quantization  of  Moduli Spaces, Springer, 2016, pp. 109--164.

\bibitem{turaev-virelizier} Vladimir Turaev and Alexis Virelizier, {\it Monoidal Categories and Topological Field Theory}, Progress in Mathematics, volume 322, Birkh\"auser (2017)



\end{thebibliography}
\end{document}